\documentclass[a4paper,11pt]{article}
\usepackage{jinstpub} 
\usepackage{lineno}

\usepackage{color}
\usepackage{ulem}
\newcommand{\Add}[1]{#1}
\newcommand{\Erase}[1]{}
\newcommand{\labroku}{LaB$_6$\ }

\newcommand{\te}{$T_{\rm e}$}

\newcommand{\dense}{$n_{\rm e}$}

\newcommand{\figs}{Fig.\,}
\newcommand{\figl}{Figure\,}
\newcommand{\degcel}{${}^{\circ}{\rm C}$}
\newcommand{\degang}{${}^{\circ}$}
\newcommand{\eqs}{Eq.\,}
\newcommand{\nitrogen}{${\rm N}_2$}

\title{\boldmath Development of the Thomson scattering measurement system for cascade arc device with indirectly heated hollow cathode}

\author[a,1]{K. Yamasaki, \note{Corresponding author}}
\author[a,2]{K. Okuda, J. Kono, A. Saito, D. Mori, R. Suzuki, Y. Kambara, R. Hamada, S. Namba}
\author[b]{K. Tomita, Y. Pan}
\author[c]{N. Tamura, C. Suzuki}
\author[d]{H. Okuno}
\affiliation[a]{Graduate School of Advanced Science and Engineering, Hiroshima University,\\
1-4-1, Kagamiyama, Higashihiroshima, Hiroshima 739-8527, Japan}
\affiliation[b]{Division of Quantum Science and Engineering, Graduate School of Engineering, Hokkaido University, \\
Kita 13, Nishi 8, Kita‑Ku, Sapporo, Hokkaido 060‑8628, Japan}
\affiliation[c]{Department of Research, National Institute for Fusion Science, \\
322-6 Oroshi-cho, Toki, Gifu 509-5292, Japan}
\affiliation[d]{Nishina Center for Accelerator-Based Science, RIKEN, \\
Wako, Saitama 351-0198, Japan}

\emailAdd{kotaro-yamasaki@hiroshima-u.ac.jp}

\abstract{
We have developed a Thomson scattering measurement system for the cascade arc discharge device designed for the plasma window (PW) application study.
The PW is one of the plasma application techniques that sustain the steep pressure gradient between high pressure (10-100 kPa) and a vacuum environment due to the thermal energy of the plasma.
Since the plasma thermal energy is the essential parameter for the pressure separation capability of PW, we installed the Thomson scattering measurement system to observe the electron density and temperature within the anode and cathode of the PW for the detailed analysis of the pressure separation capability.
The frequency-doubled Nd:YAG laser (532 nm, 200 mJ, 8 ns) was employed for the probe laser.
The scattered light was fed to the triple grating spectrometer.
The notch filter between the first and second grating eliminated the stray light, realizing a sufficiently high signal-to-noise ratio.
The Thomson scattering measurement system successfully obtained the electron density and temperature of the cascade arc plasma at 20 mm downstream from the tip of the cathode.
The installed system successfully obtained the Thomson scattering spectrum and showed that the electron density increased from $2 \times 10^{19} \, {\rm m}^{-3}$ to $7 \times 10^{19} \, {\rm m}^{-3}$ with the discharge power, while the electron temperature was almost constant at about 2 eV.
The obtained data successfully contributed to the study of the pressure separation capability of the PW.
}


\begin{document}
\maketitle
\flushbottom

\section{Introduction}
\label{sec:intro}

The plasma window (PW) is one of the plasma application techniques that sustains the pressure difference between the high-pressure (10-100 kPa) and vacuum ($\sim$ 1 Pa) environment without using solid materials such as glass and stainless steel.
The plasma inside the channel of the PW heats the neutral gas, increasing the gas temperature and reducing the flow conductance.
The flow inside the channel is considerably suppressed due to the decrease in the conductance, and the pressure difference between the gas inlet and outlet of the PW channel is kept to be substantially high\cite{Hershcovitch1998AApplications, Huang2014QuantitativeInterface, Namba2018High-densityInterfaces}.
These features enable the PW to transmit the quantum beams such as the soft-X rays, electron, and ion beams into the atmospheric pressure side without significant reduction in intensity\cite{Hershcovitch1998AApplications, MartiFelix.Heavyionstrippers.LINAC2012FR1A012012., Ikoma2019DemonstrationApplications}.
This prominent feature of PW is expected to produce new applications of quantum beam science.

One of the applications of the PW under consideration is an alternative differential pumping for helium (He) gas cell stripper in the heavy ion beam accelerators at RIKEN\cite{MartiFelix.Heavyionstrippers.LINAC2012FR1A012012.}.
The helium gas stripper must sustain the pressure of a chamber filled with helium gas up to 7 kPa and $10^{-6}$ Pa at the beamline through a channel whose diameter is larger than that of the heavy ion beams, typically 6 mm.
Ikoma {\it et al.}\cite{Ikoma2019DemonstrationApplications} developed the PW with an inner diameter of 10 mm, successfully separating 20 kPa and 0.2 Pa under the gas flow rate of 20 L/min.
However, the PWs previously developed have used needle-shaped cathodes, which frequently wear out after long hours of discharge due to the heat load\cite{LaJoie2020ViabilityStripper}.
The cathode structure, therefore, must be improved so as for PW to operate sufficiently long duration.

A hollow cathode\cite{Delcroix1974HollowArcs} can be an alternative to the conventional PW cathode.
Since the hollow cathode discharge is initiated from the inner surface of the electrode, the hollow cathode provides a large effective surface area.
This feature helps reduce the heat load at a unit area and enhance the lifetime of the cathode materials.
Also, the primary feature of the hollow cathode is its internal plasma column (IPC), a high-density and high-temperature plasma generated within the hollow electrode\cite{Delcroix1974HollowArcs}.
The IPC is suitable for the PW application because the high-density and high-temperature plasma heats the inlet gas and contributes to the pressure separation owing to the increase in the gas viscosity.
Since it has been revealed that the IPC can be created in a high-pressure environment by externally heating the hollow cathode\cite{Delcroix1974HollowArcs}, it is suggested that the PW aiming at operating in a high-pressure environment for a sufficiently long time requires an auxiliary heating system similar to those installed in hall thrusters\cite{Goebel2010CompactCathode, Becatti2017LifeThruster, Goebel2014High-CurrentThrusters} and high-density plasma source\cite{Goebel1978LanthanumProduction}.
Therefore, we have developed an 8 mm inner-diameter PW apparatus with a hollow-shaped cathode made of thermionic material (\labroku \!\!) heated indirectly.
\Erase{In this study, we installed the}\Add{The} Thomson scattering measurement system \Add{has been installed} to observe the electron density and temperature at the cathode exit to quantitatively understand the relation between the performance of the PW and the plasma thermal energy.
\Add{
In this paper, we describe the detail of the Thomson scattering measurement system, such as the hardware setting, the calibration method, and the assessment result of the accuracy of the obtained electron density and temperature.
}

\section{Experimental setup}
\label{sec:expsetup}

\begin{figure}
    \centering
    \includegraphics[keepaspectratio, scale=0.07]{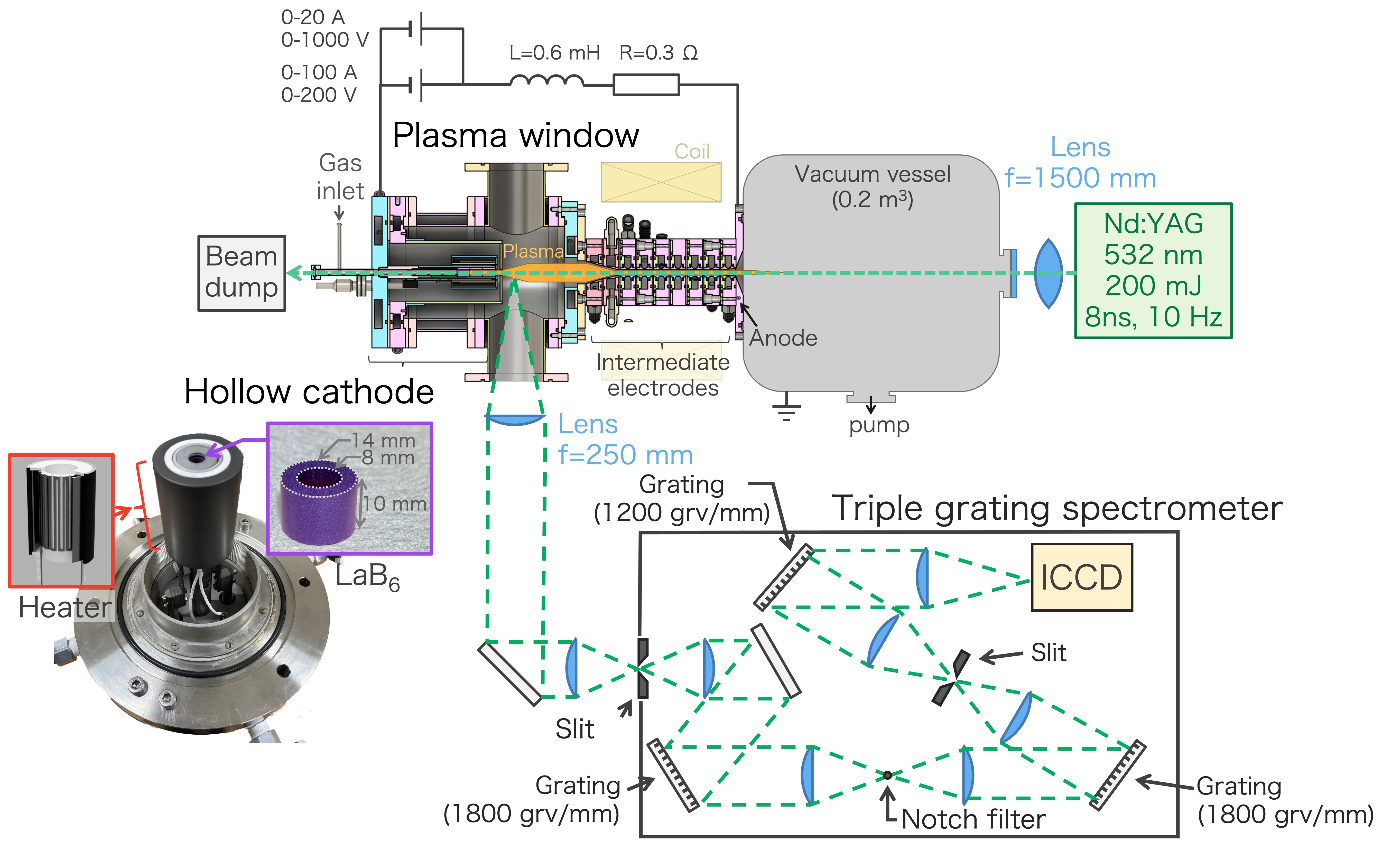}
    \caption{A schematic diagram of the plasma window, hollow cathode, vacuum vessel, and discharge circuit, together with that of the Thomson scattering measurement system. The Nd:YAG laser is focused by the lens with focal length of 1500 mm and passes through the cannel of the PW with inner diameter of 8 mm. The lens used in the triple grating spectrometer are plano-convex lens with a focal length of 250 mm.}
    \label{fig:fig02hardwarethomson}
\end{figure}

The schematic diagram of the hollow cathode and the PW itself are shown in \figs \ref{fig:fig02hardwarethomson}.
The cathode component consisted of the cylindrical thermionic material, a lanthanum hexaboride (\labroku \!\!, work function: 2.66 eV\cite{Lafferty1951BorideCathodes}), a heater, and a keeper electrode.
The length, outer and inner diameters of the \labroku cylinder were 10 mm, 14 mm, and 8 mm, respectively, and six pieces of these were inserted into the molybdenum (Mo) shaft, resulting in the inner surface area of 150 ${\rm mm}^{2}$, 6.5 times larger than that of the needle-shaped electrode previously used for the PW cathode\cite{Namba2018High-densityInterfaces}.
The large internal surface area facilitated the thermionic electron current up to 200 A when the \labroku was heated to 1700 \degcel.
The C/C (Carbon Fiber Reinforced Carbon) composite heater surrounding the Mo shaft heated the \labroku cathode.
The intermediate electrode was spaced 97 mm from the keeper electrode to perform Thomson scattering measurement downstream of the cathode exit, as shown in \figs \ref{fig:fig02hardwarethomson}.
The He gas was fed from the gas inlet tube installed in the cathode flange (see \figs \ref{fig:fig02hardwarethomson}) using a mass flow controller, whose maximum gas flow rate was 2.0 L/min.
The discharge was initiated by the ignition power supply, whose maximum supply voltage and current were 1 kV and 20 A, respectively.
The other power supply sustained the discharge, which supplied the voltage and current up to 200 V and 100 A, respectively.
The further details of the discharge device can be found in another publication\cite{Yamasaki2023DevelopmentCathode}.



\figl \ref{fig:fig02hardwarethomson} shows the schematic diagram of the Thomson scattering measurement system.
The frequency-doubled Nd:YAG laser (532 nm, 200 mJ, 8 ns) was employed as the probe laser and was fed from the window at the end of the vacuum vessel.
The laser was focused at the cathode exit using the lens with a focal length of 1500 mm, which was located close to the window on the atmospheric side.
The 90\degang \ scattered light at 20 mm downstream from the keeper electrode was introduced to the triple grating spectrometer using a pair of plano-convex lenses whose focal length was 250 mm.
The triple grating spectrometer consisted of three gratings and a notch filter.
The first and second grating (both 1800 grooves/mm) dispersed and inversely dispersed the incident light to eliminate the stray light from the out of the region of interest\cite{KevinFinch2020AAnalysis, Kaloyan2022FirstDevice, Yamamoto2012MeasurementTechnique, Tomita2017MeasurementPANTA}.
The notch filter between the two gratings occluded the stray light spectrum around 532 nm to improve the signal-to-noise ratio.
The light passed through the second slit was dispersed by the third grating (1200 groove/mm), and the spectrum was observed by an ICCD camera (Princeton Instruments PI-MAX1KUV-18-43-U, 1024×1024 pix, 13 $\mu$m pitch).
The 1st and 2nd slit width was set to 200 $\mu$m for this study.
The sensitivity of the system to the electron density and the inverse linear dispersion (ILD) was obtained by the Rayleigh and Raman spectrum from \nitrogen \ molecules under the nitrogen gas filling pressure range of 2-16 kPa, respectively.
\section{Results and discussion}
\label{sec:resdis}

\paragraph{Beam focusing condition}

\begin{figure}
        \centering
        \includegraphics[keepaspectratio, scale=1.2]{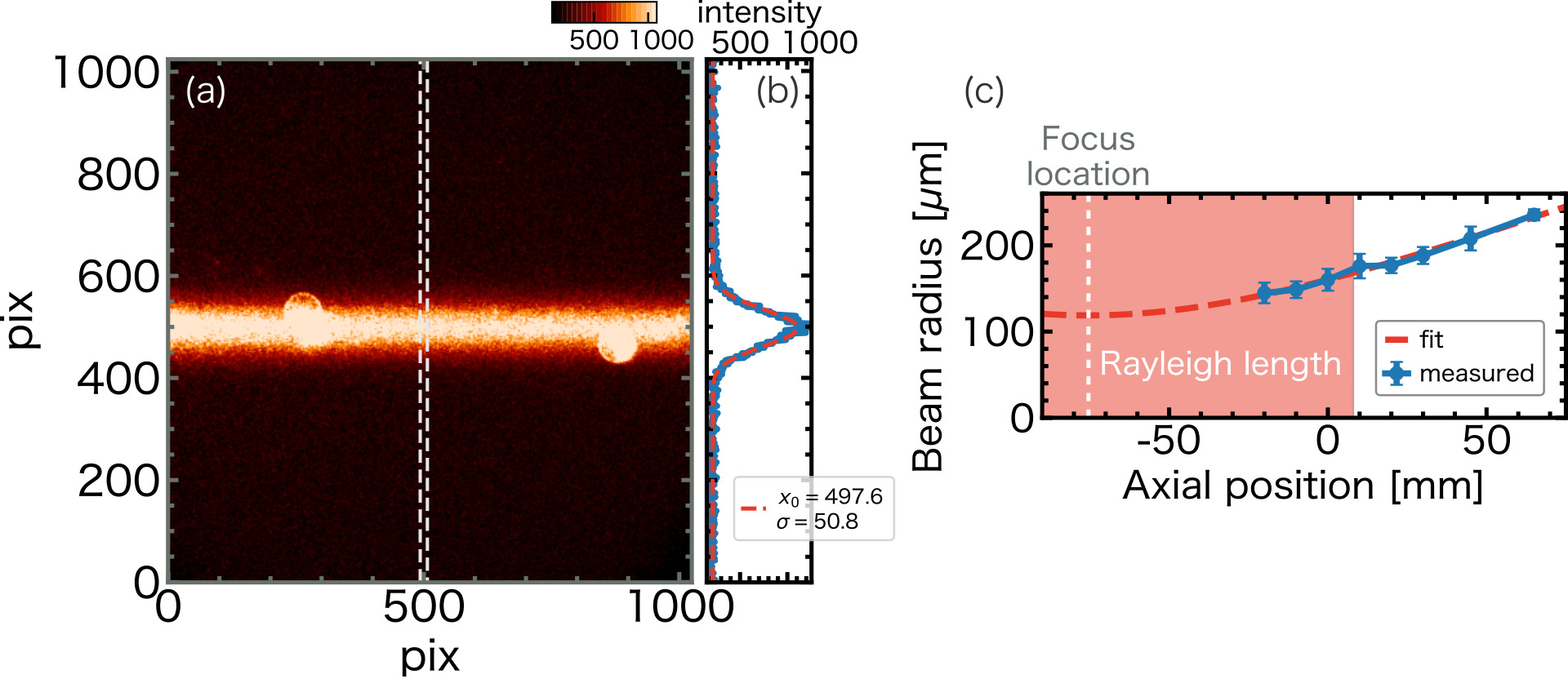}
        \caption{(a) The typical image of Rayleigh scattering along the beam path and (b) one dimensional profile of the intensity across the beam path. The imaging of the Rayleigh scattering was performed at the atmospheric pressure condition with the observation angle of 90 degrees from the beam path. The dots on the left and right of the image represent the Mie scattering light from the dust. The one-dimensional image was binned by fifteen pixels in horizontal direction. (c) The axial profile of the beam radius along the beam path. The Thomson scattering measurement was performed at axial position of 0 mm.}
        \label{fig:fig03RayleighBeamWidth}
\end{figure}

The beam radius was measured by the image of Rayleigh scattering light at atmospheric pressure.
The Rayleigh scattering light was observed 90\degang \ from the beam path and was collected by a lens with a focal length of 100 mm and ICCD, resulting in an image magnification of about 3.9.
The typical image of the Rayleigh scattering light is shown in \figs \ref{fig:fig03RayleighBeamWidth}, manifesting the Gaussian distribution of the laser intensity (see \figs \ref{fig:fig03RayleighBeamWidth} (b)).
The dots in the left and right of \figs \ref{fig:fig03RayleighBeamWidth}(a) represent the Mie scattering light from the dust.
The 1/e radius of the Rayleigh scattering light intensity was chosen as the beam radius in this study.
\figl \ref{fig:fig03RayleighBeamWidth} (c) shows the axial profile of the observed beam radius and the fitted curve of the axial profile of the Gaussian beam radius defined as follows,
\begin{eqnarray}
    w(z; w_0, z_0, \lambda)
    = w_0
    \left[
    1 + 
    \left\{
    \frac{\lambda (z-z_0)}{\pi w_0^2}
    \right\}^2
    \right]^{\frac{1}{2}}.
\end{eqnarray}
Here, $w_0, \lambda, z$ and $z_0$ represent the beam radius at the waist, the wavelength of the laser, axial position along the beam path referenced from the axial position where the Thomson scattering measurement was performed, and the axial position of the beam waist, respectively.
The fitting result shows that the beam was focused 75 mm downstream of the observation point, namely 95 mm away from the keeper electrode, resulting in the beam radius at the observation point of 160 $\mu$m.
Also, the fitting analysis shows that the observation point is barely within the Rayleigh length from the focus, suggesting that the probe beam is nicely focused.

\paragraph{Calibration}

The density calibration was performed by the Rayleigh scattering signal with various \nitrogen \ filling gas pressure conditions in the range of 2-16 kPa.
In this case, a hundred shots were accumulated for collecting the Rayleigh scattering signal.
The inclination of the Rayleigh intensity to the \nitrogen \ density was $2.10 \times 10^{-22} \ {\rm count} \, {\rm m}^{3}$, resulting in the sensitivity of the Thomson scattering signal of $3.66 \times 10^{19} \ {\rm m}^{-3}/{\rm count}/{\rm shot}$.
The region masked by the notch filter was confirmed by observing the spectrum of the white light source.
In this study, we selected the horizontal pixel from 450 to 520 as the masked region. 

\begin{figure}
\begin{tabular}{clr}
   \begin{minipage}[t]{0.53\hsize}
        \centering
        \includegraphics[keepaspectratio, scale=1.15]{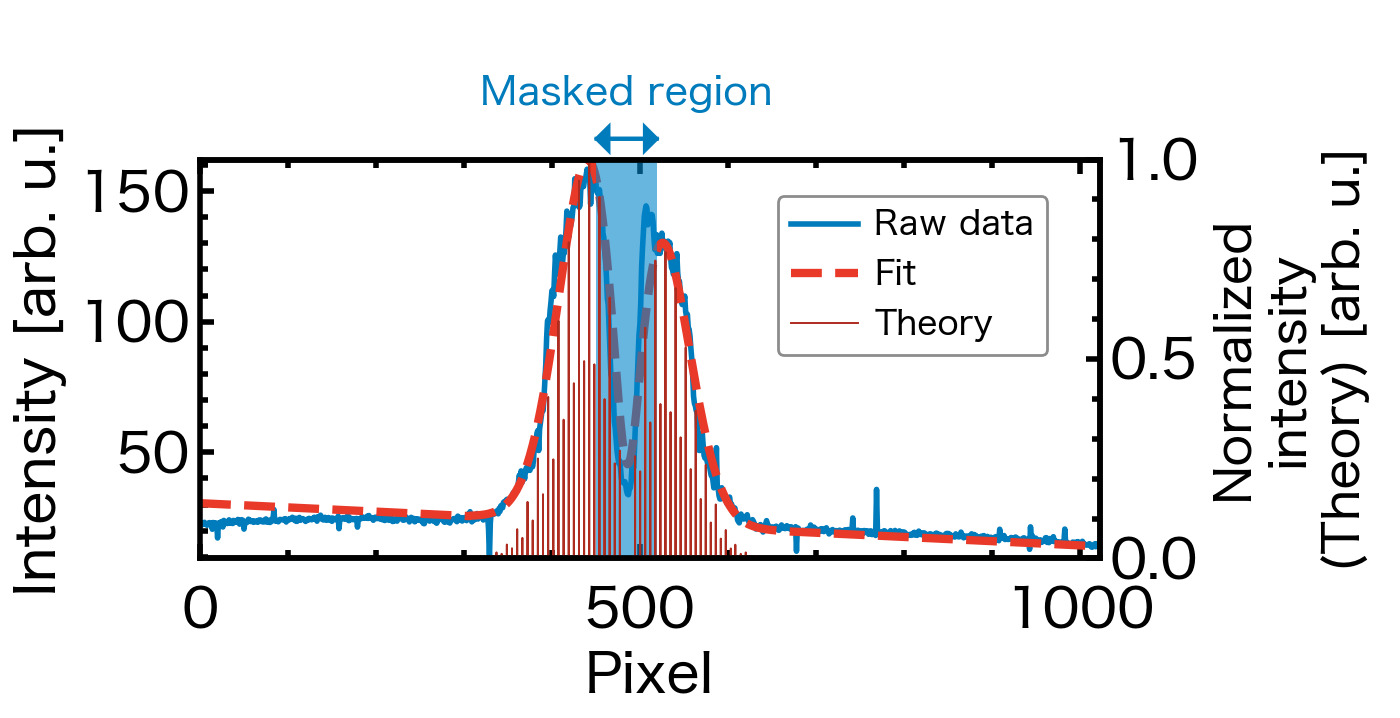}
        \caption{The raw signal of the Raman spectrum together with the fitting result of $I_{\rm Raman,obs}(\lambda;A,\sigma,a,b,r_{ILD})$ to determine the inverse linear dispersion (ILD) of the spectroscopic system. The thin red line represents the theoretically obtained Raman spectrum $I_{\rm Raman}(\lambda)$. The blue-hatched region represents the range where the notch filter occludes the spectrum signal.}
        \label{fig:fig07RamanCalib}
    \end{minipage} & &
    \begin{minipage}[t]{0.45\hsize}
        \centering
        \includegraphics[keepaspectratio, scale=1.15]{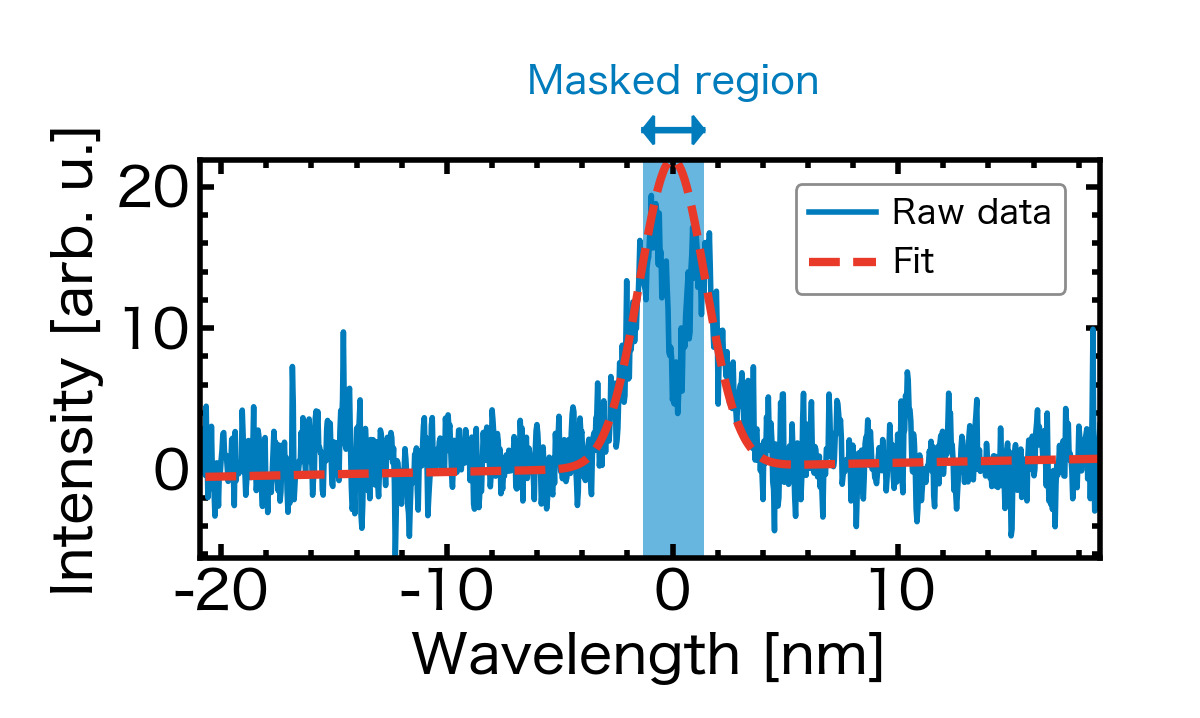}
        \caption{The Thomson scattering spectrum obtained under the condition with the gas flow rate of 0.07 L/min and discharge current of 100 A together with the fitting result. The blue-hatched region represents the range where the notch filter occludes the spectrum signal.}
        \label{fig:fig08ThomsonExample}
    \end{minipage}
\end{tabular}
\end{figure}

By inserting the notch filter, we can obtain the Raman spectrum around the laser wavelength.
The wavelength and relative intensity of each Raman spectrum can be theoretically obtained, as shown in the thin red line in \figs \ref{fig:fig07RamanCalib}; thus, the ILD of the triple grating spectrometer can be calculated from the Raman spectrum.
Since the Raman spectrum $I_{\rm Raman}(\lambda)$ is distorted by the instrumental function of the spectrometer, the observed spectrum can be described by the convolution of the Raman spectrum and the instrumental function as follows,
\begin{eqnarray}
    I_{\rm Raman,obs}(\lambda; A, \sigma, a, b, r_{ILD})
    &=& A
    \int_{-\infty}^{\infty}
      I_{\rm Raman}(\lambda')
      f(\lambda-\lambda', \sigma, r_{ILD})
    d\lambda' + a \lambda + b \label{eq:ramanfunc} \\
    f(\lambda-\lambda', \sigma, r_{ILD})
    &=&
    \frac{1}{\sqrt{\pi}r_{ILD}\sigma}
    {\rm exp}
    \left\{
      - \left(
        \frac{\lambda-\lambda'}{r_{ILD}\sigma}
      \right)^2
    \right\}.
\end{eqnarray}
Here, $\lambda, I_{\rm Raman,obs}(\lambda; A, \sigma, a, b, r_{ILD}), \sigma, f(\lambda-\lambda', \sigma, r_{ILD}), a, b$ and $r_{ILD}$ represent the wavelength, the observed Raman spectrum, the 1/e half width of the instrumental function in the unit of a pixel, the instrumental function, the inclination and offset of the background component, and the ILD, respectively.
In this study, we assumed that the shape of the Rayleigh scattering spectrum represents the instrumental function of the spectroscopic system.
Therefore, we assumed the instrumental function as the Gaussian function, whose 1/e width is the same as that of the Rayleigh scattering spectrum.
\figl \ref{fig:fig07RamanCalib} shows the raw signal of the Raman spectrum, taken with the \nitrogen \ filling pressure of 10 kPa, and the fitting function, showing that \eqs (\ref{eq:ramanfunc}) is a good approximation of the observed Raman spectrum.
In this case, the ILD was 0.039 nm/pixel, corresponding to 3.0 nm/mm at the ICCD detector.
The obtained ILD and the notch filter spectrum suggest that the width of the masked region corresponds to 2.73 nm, namely 530.7-533.4 nm.

\paragraph{Thomson scattering measurement results, the accuracy, and implication}

\begin{figure}
\begin{tabular}{ll}
   \begin{minipage}[t]{0.41\hsize}
        \centering
        \includegraphics[keepaspectratio, scale=0.87]{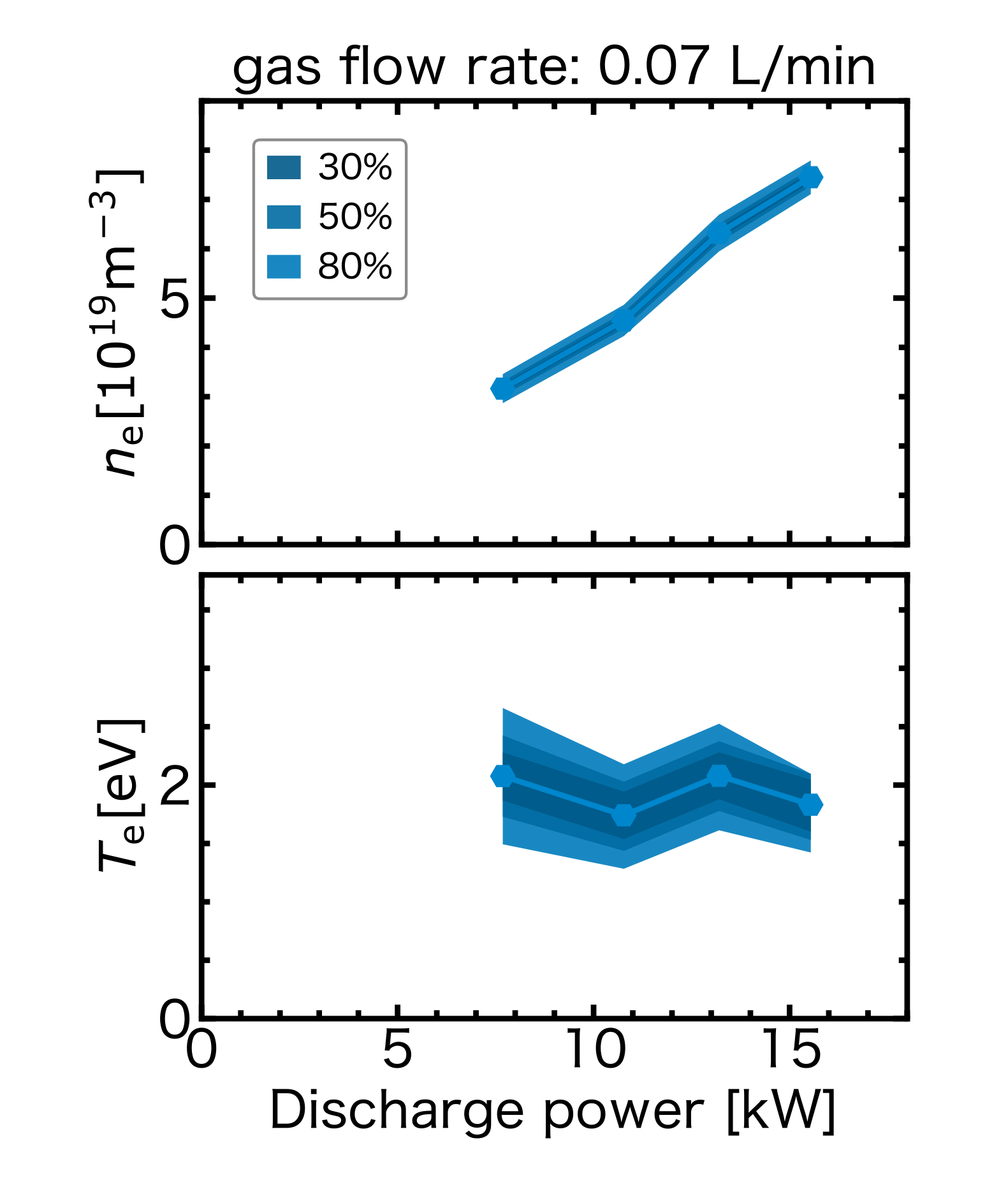}
        \caption{The dependence of (a) \dense \ and (b) \te \ on the discharge power for the case with the gas flow rate of 0.07 L/min. The hatched regions represent the range where 30\%, 50\%, and 80\% of the ensemble of fitting results are located, indicating the accuracy of the fitting results.}
        \label{fig:fig09TeNeTrend}
    \end{minipage} &
    \begin{minipage}[t]{0.55\hsize}
        \centering
        \includegraphics[keepaspectratio, scale=0.9]{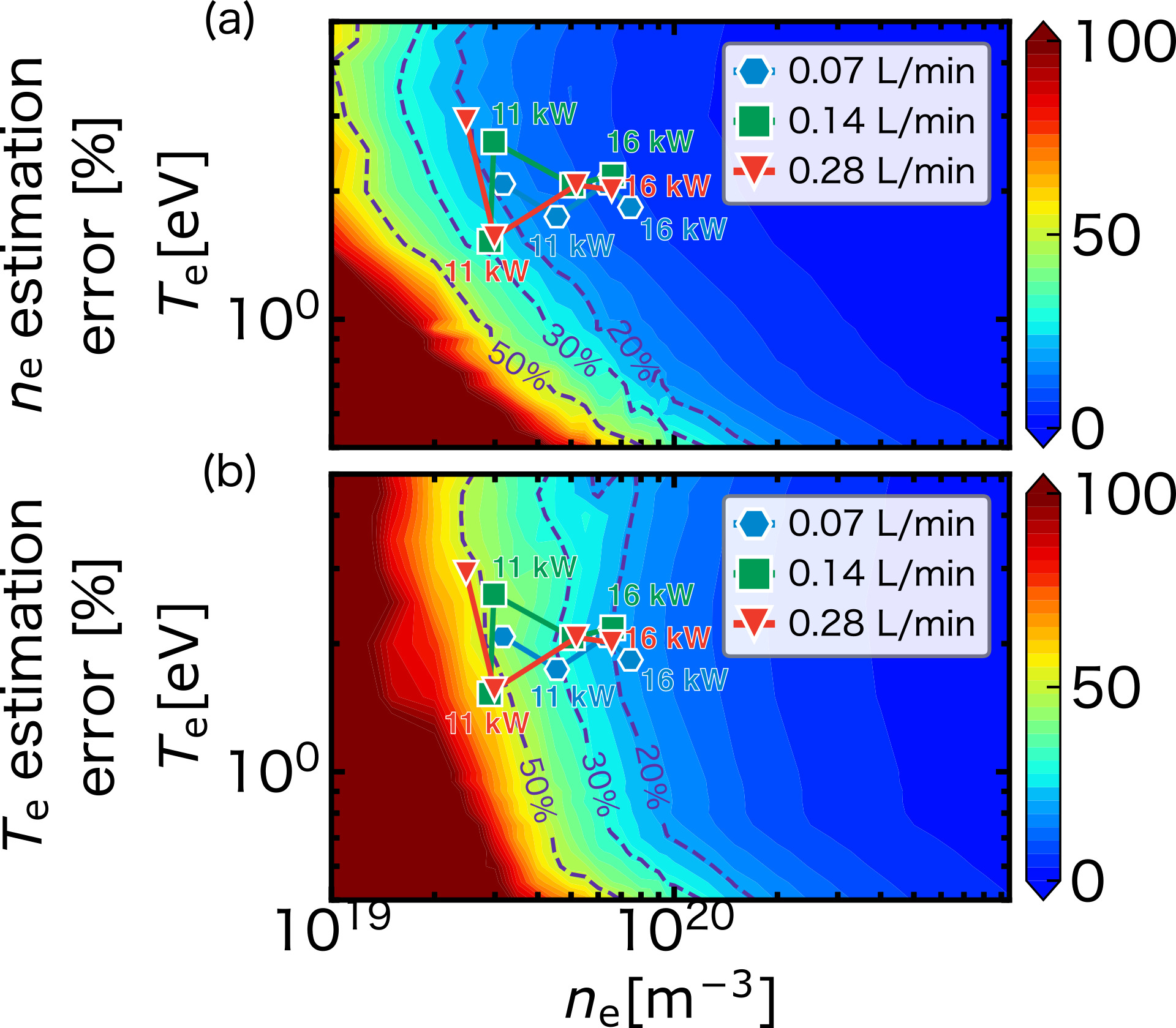}
        \caption{The dependence of the estimation error of (a) \dense \ and (b) \te. In this case, the width of the range, where 80\% of the estimation ensembles nearest from the designated electron density or temperature are located, is adopted as the error of the estimation. The color contours show the ratio of the error to the true value.}
        \label{fig:fig10ErrorEstimate}
    \end{minipage}
\end{tabular}
\end{figure}
\figl \ref{fig:fig08ThomsonExample} shows the obtained Thomson scattering spectrum under the discharge condition with the gas flow rate of 0.07 L/min and discharge current of 100 A.
The dipped region around the wavelength center represents the occlusion of the spectrum by the notch filter. Around the masked region, we can see the tail of the Thomson scattering spectrum.
To obtain the electron density and temperature, we fit the Gaussian function with the following form,
\begin{eqnarray}
    I_T(\lambda; A_T, \sigma_T, a_T, b_T) =
    A_T {\rm exp}
    \left\{
      - \left(
        \frac{\lambda}{\sigma_T}
      \right)^2
    \right\}
    + a_T \lambda + b_T.
\end{eqnarray}
Here, $A_T, \sigma_T, a_T$ and $b_T$ represent the height and 1/e half width of the Gaussian, the inclination and offset of the background component, respectively.
The Gaussian fitting is performed to the data from -12.2 nm to 11.2 nm except for the masked region from -1.33 nm to 1.40 nm.
The electron density \dense \ and temperature \te \ are calculated as follows,
\begin{eqnarray}
    n_{\rm e} &=& 3.66 \times 10^{19} \frac{\sqrt{\pi}A_T \sigma_T}{N_{\rm shot}} \ [{\rm m}^{-3}] \\
    T_{\rm e} &=& \frac{m_{\rm e}}{2{\rm e}}
    \left\{
      \frac{\sqrt{\sigma_T^2 - \sigma_{\rm inst}^2}}{2 {\rm sin}\left( \frac{\theta}{2} \right)}
      \frac{c}{\lambda_{\rm laser}}
    \right\}^2 \ [{\rm eV}].
\end{eqnarray}
Here, $N_{\rm shot}, \sigma_{\rm inst}, \theta, \lambda_{\rm laser}$ and $c$ represent the number of accumulations for obtaining the Thomson scattering signal, the 1/e half width of the instrumental function in the unit of wavelength, the observation angle from the laser path, the wavelength of the laser, and the speed of light, respectively.
The dependence of the observed \dense \ and \te \ on the discharge power for the case with the gas flow rate of 0.07 L/min is shown in \figs \ref{fig:fig09TeNeTrend}, suggesting that the electron density monotonically increases with the discharge power, which is also confirmed by the Stark broadening measurement of the impurity hydrogen line, while the electron temperature seems to be irrelevant.
Similar trend was found for the cases with the gas flow rate of 0.14 and 0.28 L/min.
The error of the electron temperature and density \Add{due to the fitting stability against the noise and the available spectral range due to the notch filter} are estimated by performing Gaussian fitting to the test data defined as follows,
\begin{eqnarray}
    I_{\rm test}(\lambda) =
    I_T(\lambda; A_T(n_{\rm e}, T_{\rm e}), \sigma_T(T_{\rm e}), a_T, b_T) +
    \delta(\lambda; \sigma_{\rm noise})
\end{eqnarray}
Here, $A_T(n_{\rm e}, T_{\rm e})$ and $\sigma_T(T_{\rm e})$ represent the height and 1/e half width of the Gaussian profile corresponding to the designated electron density and temperature, respectively.
The noise component $\delta(\lambda; \sigma_{\rm noise})$ was produced to be the normal distribution with a standard deviation $\sigma_{\rm noise}$ same as that of the noise outside the fitting region, the observed noise in the area whose wavelength is less than -12.2 nm or more than 11.2 nm.
In this study, 3000 sets of noise data are produced to estimate the accuracy of the fitting result.
The accuracy of the fitting result is evaluated as the range of the electron density or temperature, where 30\%, 50\%, and 80\% of the ensembles nearest from the designated \dense \ and \te \ are located.
The step by step process for the accuracy estimation is as follows;
\begin{enumerate}
    \setlength\baselineskip{3pt}
    \setlength{\itemsep}{-3pt}
    \item Calculate \dense \ and \te \ for each test data with different noise.
    \item Calculate the absolute value of difference between the designated values and the ensembles.
    \label{proc:diff}
    \item Sort the ensemble in ascending order of the difference calculated in step \ref{proc:diff}.
    \item Gather 30\%, 50\%, and 80\% of the ensemble from the beginning of the sorted ensemble.
    \label{proc:gather}
    \item Find the minimum and maximum of \dense \ and \te \ for each set created in step \ref{proc:gather}, which define the lower and upper bound of the region, respectively.
\end{enumerate}
\noindent
The estimate should be more accurate when the range of the ensemble is narrower.
Therefore, \figs \ref{fig:fig09TeNeTrend} suggests that the accuracy of the electron temperature is high in the cases with higher discharge power.
In contrast, the accuracy is relatively poor with lower discharge power cases.
\figl \ref{fig:fig10ErrorEstimate} shows the dependence of the electron density and temperature estimation error on the electron density and temperature itself.
In this case, the width of the range, where 80\% of the estimation ensembles nearest to the designated electron density or temperature are located, is chosen as the error or the estimation.
\figl \ref{fig:fig10ErrorEstimate} (a) and (b) shows that the estimation error becomes smaller with larger electron density.
In this study, the electron density increased with higher discharge power.
Therefore, we can obtain the electron density and temperature more accurately when the discharge power is more than 13 kW\Add{, where the estimation error of the electron temperature is less than 25\%}.
On the other hand, the error of the estimation, especially the electron temperature, increased to about 50\% for the cases with discharge power of less than 12 kW, suggesting that signal-to-noise ratio must be improved for the detailed analysis of the data with lower discharge power conditions.
The detailed analysis on the performance of the PW, using the obtained \dense \ and \te, is described in other publication\cite{Yamasaki2023DevelopmentCathode}.


\section{Summary}
\label{sec:summary}

We have developed a Thomson scattering measurement system for the cascade arc discharge device designed for the plasma window (PW) application study.
The Thomson scattering measurement system successfully obtained the electron density and temperature of the cascade arc plasma at 20 mm downstream from the tip of the cathode.
The Thomson scattering spectrum was successfully obtained and showed that the electron density increased from $2 \times 10^{19} \, {\rm m}^{-3}$ to $7 \times 10^{19} \, {\rm m}^{-3}$ with the discharge power, while the electron temperature was almost constant at about 2 eV.
The analysis on the estimation error of the electron density and temperature showed that the current system can measure the electron density and temperature with sufficient accuracy when the discharge power is more than 13 kW.
The obtained data successfully contributed to the study of the pressure separation capability of the PW.

\acknowledgments

This research was supported by JSPS KAKENHI under Grant No. JP20H00141, O. S. G. Fund, Furukawa research fund, and NIFS Collaboration Research Program under Grant No. NIFS NIFS22KIIH003, NIFS22KIIH011, and NIFS23KIIH016.


\bibliographystyle{JHEP}
\bibliography{references}






\end{document}